\documentclass{moriond}

\def\be{\begin{equation}}
\def\ee{\end{equation}}
\def\bea{\begin{eqnarray}}
\def\eea{\end{eqnarray}}

\newcommand{\Photo}{}

\begin{document}
\vspace*{4cm}
\title{A $p_T$-ratio observable for studies of intrinsic transverse momentum of partons from Drell-Yan $p_T$ spectra}

\author{ Wenxiao Zhan, Siqi Yang, Minghui Liu, Liang Han }
	
\address{University of Science and Technology of China, Hefei, China}

\author{ Francesco Hautmann }

\address{University of Oxford, Oxford, UK}

\maketitle\abstracts{
The determination of the intrinsic transverse momentum distribution 
of partons is central both 
for applications of parton shower Monte Carlo generators and 
for QCD studies of transverse momentum dependent (TMD) parton 
densities.   Valuable information on this distribution is 
provided by experimental measurements of Drell-Yan transverse 
momentum $p_T$,   in the region of low transverse momenta, with fine 
binning in $p_T$. However, such fine-binning measurements are 
challenging, as they require 
an extremely delicate control of systematic uncertainties. 
We suggest a $p_T$ observable  based on measuring ratios between 
cross sections of
suitably defined low-$p_T$ and high-$p_T$  regions. This observable  
does not rely on any dedicated partition of bins  and has lower 
systematic uncertainties, and is shown to  provide 
a good sensitivity to the intrinsic  transverse momentum. 
}

Measurements of transverse momentum spectra of 
electroweak bosons via Drell-Yan (DY) lepton-pair production
are  used in a wide range of collider physics  studies,  
 from the determination of the 
strong coupling~\cite{ATLAS:2023lhg,Camarda:2022qdg} 
to the tuning of  
Monte Carlo event generators~\cite{CMS:2024goo},  
to the extraction   of  TMD 
 parton distributions~\cite{Angeles-Martinez:2015sea,Abdulov:2021ivr} both
in analytic-resummation~\cite{Moos:2025sal,Bacchetta:2024qre,Bury:2022czx,Hautmann:2020cyp} 
and parton-branching~\cite{Hautmann:2025fkw,Bubanja:2023nrd}
approaches. 

These studies illustrate  that high-precision 
measurements of the region of low DY transverse momenta,  
 $\Lambda_{\rm{QCD}}
{\raisebox{-.6ex}{\rlap{$\,\sim\,$}} \raisebox{.4ex}{$\,<\,$}} p_T(ll) \ll m(ll)$, 
with fine binning in $p_T (ll)$ are able to greatly increase our ability
to unravel non-perturbative dynamics in the strong-interaction sector of the 
Standard Model.
On the other hand, measurements of the DY process by the 
ATLAS~\cite{ATLAS:2019zci} and CMS~\cite{CMS:2019raw} Collaborations 
indicate that such fine-binned analyses require 
 an extremely delicate control  
of systematic uncertainties, which makes 
the determination of  the low-$p_T(ll)$ fine-binned structure
experimentally challenging.
It thus becomes important to 
devise methodologies which do not require 
the fine-binned measurement of $p_T(ll)$ distributions, 
and explore their capabilities.  

In this article we discuss one such methodology, based on 
our work published in Ref.~\cite{Zhan:2024lym}.  
We consider the process $ pp \to Z / \gamma \to l^+ l^- $, and 
examine scenarios with coarse-grained binning of 
the region  $0 < p_T(ll) < p_{T , {\rm{max}}} $ (with $p_{T , {\rm {max}}} $ small
compared to  the DY  invariant mass $ m(ll)$).  
To be definite, 
we subdivide the $p_T$ region into two bins 
with separation momentum $p_s$,
$p_T < p_s$ and $p_T > p_s$; then  we study the 
sensitivity  of the ratio between the
low-$p_T (ll)$ cross section (dominated by 
TMD dynamics and resummed soft-parton radiation) and
the high-$p_T(ll) $ cross section
 (dominated by  fixed-order hard-parton radiation), 
to the intrinsic transverse momentum $k_T$ distribution, as a 
 function of varying $p_s$. 

The basic outcome of applying this methodology 
is that a good  sensitivity  
can be achieved from the fall-off of the $p_T $-ratio 
with increasing 
 gaussian width of the intrinsic  $k_T$ distribution. 
This results into  
 significant advantages,  
from the practical viewpoint 
of controlling experimental systematic uncertainties, 
 with respect to measuring 
 the complete  fine-binned $p_T$ shape. 

To carry out this study, we use 
Monte Carlo predictions computed by 
matching, according to the 
method~\cite{BermudezMartinez:2019anj,BermudezMartinez:2020tys,Abdulhamid:2021xtt,Yang:2022qgk}, 
next-to-leading-order (NLO) hard-scattering matrix elements 
from {\scshape MadGraph5\_aMC@NLO}~\cite{Alwall:2014hca} (hereafter, MCatNLO)  
 with TMD parton distributions and showers
obtained from parton-branching evolution~\cite{Hautmann:2017xtx,Hautmann:2017fcj,BermudezMartinez:2018fsv} and implemented in the {\sc Cascade} 
event generator~\cite{CASCADE:2021bxe,CASCADE:2010clj}. This calculation has next-to-leading-logarithmic (NLL) accuracy in the 
resummation of the DY spectrum for small $p_T  (ll) / m (ll)$. The method can be extended to next-to-next-to-leading-logarithmic (NNLL) accuracy 
using the technique of Refs.~\cite{Martinez:2024mou,Martinez:2024twn}. Besides  the DY $p_T$ spectrum, the TMD parton distributions and showers used in this work have  
also been applied  to make predictions 
for DY + jets observables by means of multi-jet merging 
techniques~\cite{BermudezMartinez:2021lxz,BermudezMartinez:2021zlg,BermudezMartinez:2022bpj}.

\begin{figure}
	\begin{minipage}{0.45\linewidth}
		\centerline{\includegraphics[width=\linewidth]{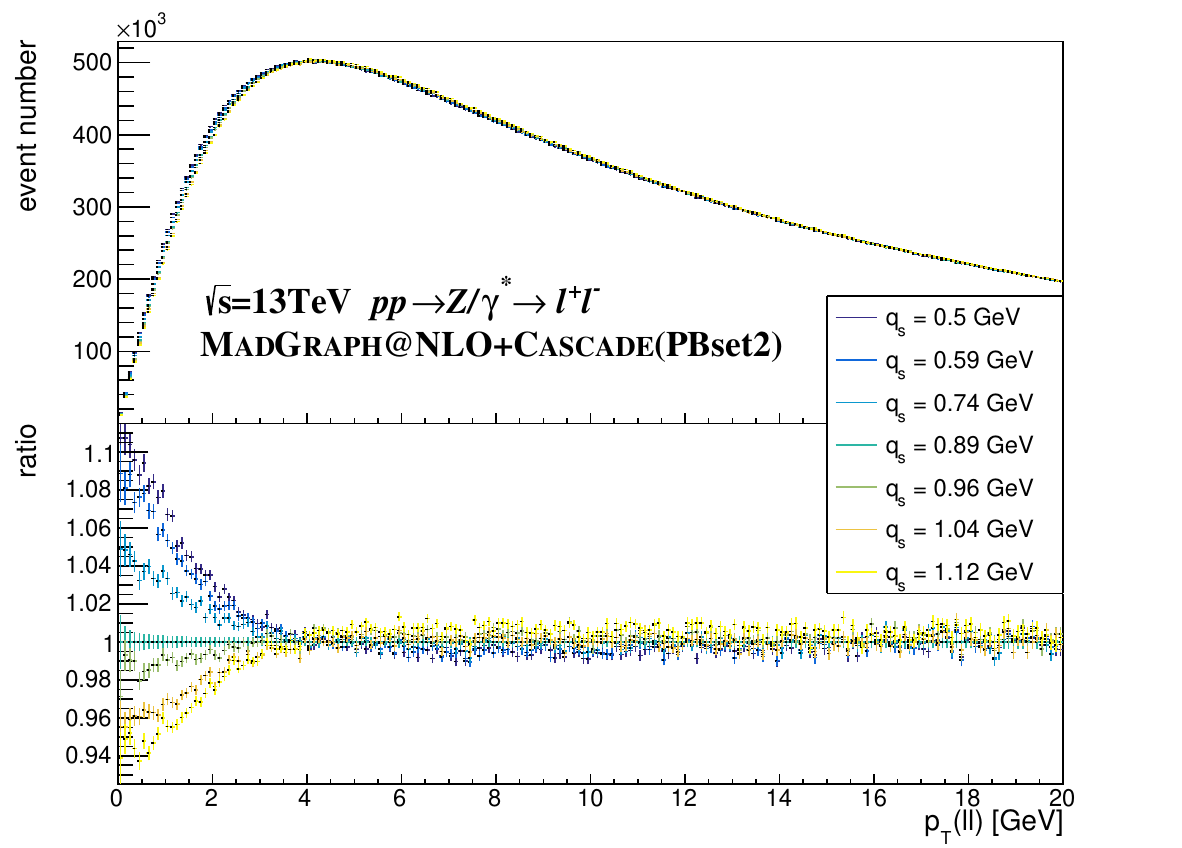}}
	\end{minipage}
	\begin{minipage}{0.47\linewidth}
		\centerline{\includegraphics[width=\linewidth]{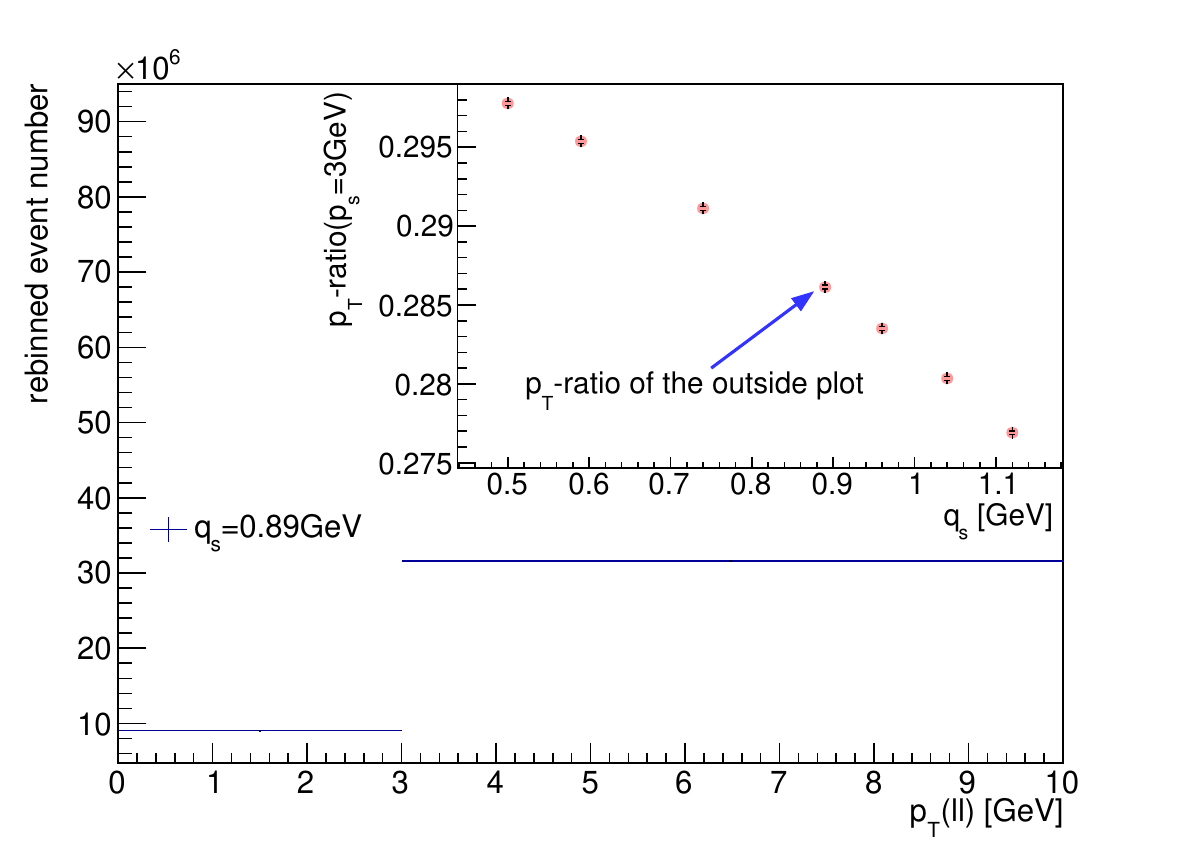}}
	\end{minipage}
	\caption{(left)     Transverse momentum distributions from
    the template samples
    with different intrinsic-$k_T$ gaussian width $q_s$ in the 
    full phase space, with the ratio plot
    at the bottom obtained  by taking the ratio to the distribution with $q_s$ = 0.89; (right) 
    the two-bin distribution of lower and higher $p_T$ and, in the inset,  the  $q_s$ dependence of  the     $p_T$-ratio, for $p_s = $ 3 GeV.}
	\label{fig:show_ratio}
\end{figure}

Using the   MCatNLO + {\sc Cascade} Monte Carlo results,  
we explore the  intrinsic-$k_T$   parameter space 
by generating template samples with
different  values    of the gaussian width $q_s$ of the 
  non-perturbative TMD 
  distribution~\cite{Bubanja:2023nrd,BermudezMartinez:2018fsv}.      
In Fig.~\ref{fig:show_ratio}~\cite{Zhan:2024lym}, on the left 
we show the $ p_T(ll)  $ distribution   
   from the template samples  over the range $ p_T(ll) <  20$ GeV;   
 on the right,   we introduce the momentum
parameter $p_s$ to separate the regions $p_T < p_s$ and $p_T > p_s$, and
construct the 2-bin $p_T$-ratio between the lower and higher $p_T$ regions, 
 defined from the integral of event numbers in the 
relatively low ($p_L$) and high ($p_H$) region, 
$$	 p_T\mathrm{-ratio} = {p_L} / {p_H} . $$

The sensitivity to the  intrinsic transverse momentum of partons 
comes primarily from the low-$p_T$ region. 
The extraction of the  intrinsic-$k_T$   parameter  $q_s$ 
from the $p_T$-ratio will  depend critically on the 
 separation momentum $p_s$. With too low a value of 
  $p_s$,  sensitivity may be lost due to 
soft-gluon splitting transverse momenta moving to 
 the upper region; with too high a value of 
  $p_s$,  the information on low $p_T$ may become undetectable.

\begin{figure}
	\begin{minipage}{0.47\linewidth}
		\centerline{\includegraphics[width=\linewidth]{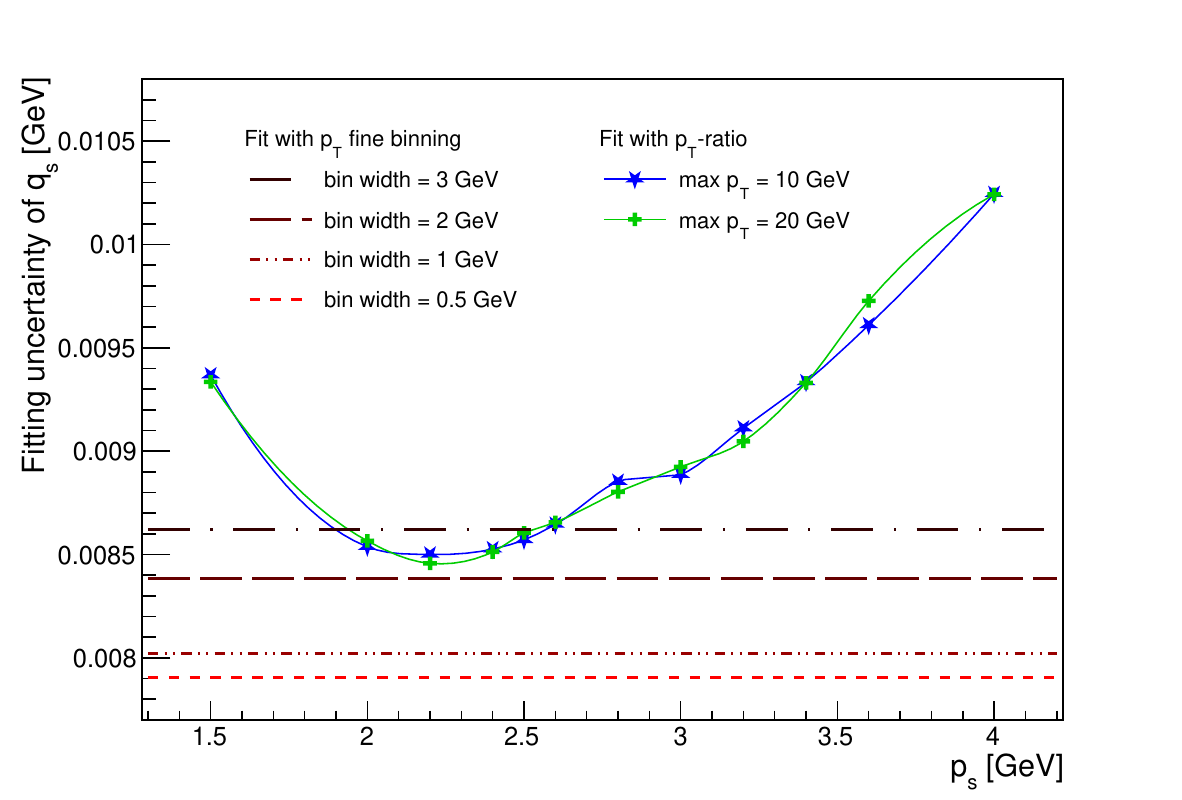}}
	\end{minipage}
	\begin{minipage}{0.47\linewidth}
		\centerline{\includegraphics[width=\linewidth]{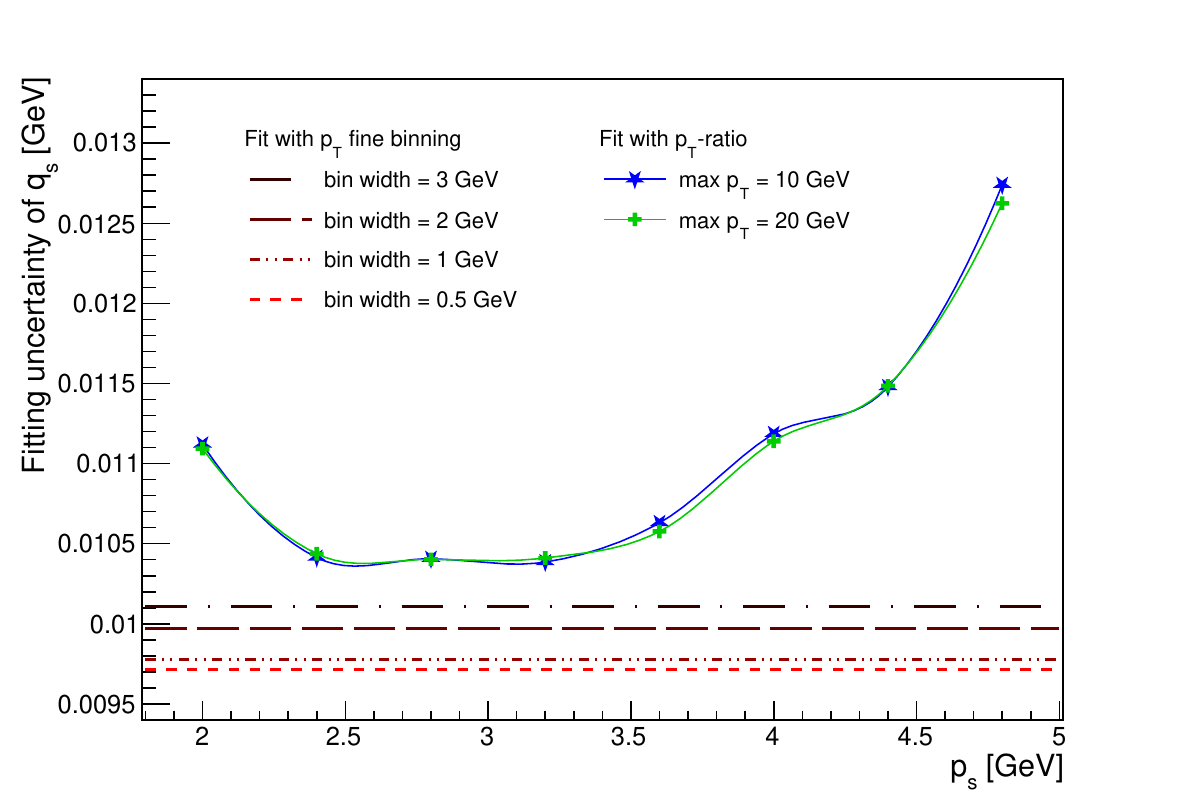}}
	\end{minipage}
	\caption{Statistical uncertainties from binned-$p_T$ and $p_T$-ratio in the test with (left) and without (right) migration effect. }
	\label{fig:ss}
\end{figure}

A sensitivity test is performed in 
Fig.~\ref{fig:ss}~\cite{Zhan:2024lym} 
for the cases of  both the fine-binned $p_T$ and the $p_T$-ratio. 
The sensitivity is represented by the fitting uncertainties 
of $q_s$ in the two cases, with the fits being performed as 
described in   Ref.~\cite{Zhan:2024lym}.  In the case of 
binned $p_T$, we apply different bin widths from 0.5 GeV to 3 GeV.  
In the case of the $p_T$-ratio, we examine the sensitivity change for 
different choices of $p_s$. 
The  extraction of  $q_s$ is carried out  using  
the template samples  in the phase space 
with final-state lepton selections $p_T(l^\pm)>25$ GeV and $|\eta(l^\pm)|<2.4$, 
in which leptons are dressed with photons reconstructed within 
$\Delta R=\sqrt{(\Delta\eta)^2+(\Delta\phi)^2}<0.1$. This is close to 
the fiducial phase space in real experiments at ATLAS and CMS. 
Also, in real experiments 
momentum resolution causes bin-to-bin migration. Since the migration effect is a dominant systematic uncertainty, it is essential to verify that such an effect does not spoil 
the performance capabilities of  the $p_T$-ratio. To 
study this, we apply a 3\% resolution on the four-momentum of the dressed leptons. 
The left panel of Fig.~\ref{fig:ss} shows the results of the sensitivity test 
without migration effect; the right panel shows the results after inclusion of the 
migration effect. 
The overall conclusion from this test is 
that the $p_T$-ratio still has a good sensitivity to the intrinsic $k_T$, so that this methodology can be applied directly in the experiments for TMD performance 
studies.

Given the difficulties in achieving  finer binning in the very low $p_T(ll)$ range,
we next test the $p_T$-ratio methodology, based on  
coarse-grained  binning in the  low-$p_T$ fiducial region,  by applying it to real
experimental data.  The CMS collaboration has measured the $p_T(ll)$ distribution in a wide mass range from 50 GeV to 1 TeV~\cite{CMS:2022ubq}. This measurement has been used  for the determination of the $q_s$  dependence on DY mass in 
Ref.~\cite{Bubanja:2023nrd}. We here perform an extraction of  $q_s$ from this measurement  using the  $p_T$-ratio. Besides testing the feasibility of
the $p_T$-ratio proposal with real data, we also aim to check that the  $p_T$-ratio is not biased by the high-$p_T$ region, where contributions from higher jet
multiplicities are important~\cite{BermudezMartinez:2021lxz}.

\begin{figure}
	\centerline{\includegraphics[width=0.5\linewidth]{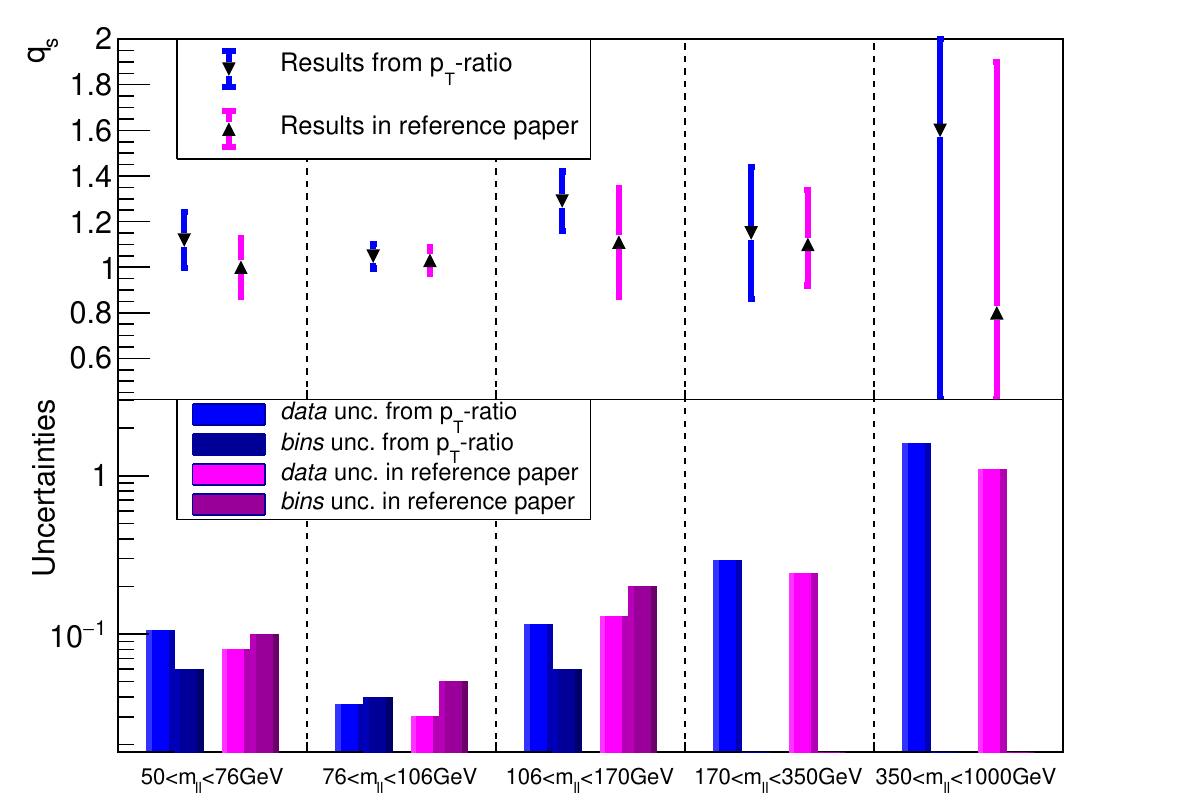}}
	\caption{Determination of intrinsic $k_T$ parameter  $q_s$. The upper panel contains the $q_s$ values extracted independently in different $m_{DY}$ bins, compared with the results in the reference paper~\protect\cite{Bubanja:2023nrd}. The error bars represent the combinations of the corresponding uncertainties in the lower pad. The lower panel contains a comparison of  each source of uncertainties. In the last two bins we only consider \textit{data} uncertainties.}
	\label{fig:result}
\end{figure}

Results are reported in Fig.~\ref{fig:result}~\cite{Zhan:2024lym}. 
Statistical uncertainties and scale uncertainties 
are considered, and compared to those in the 
analysis of Ref.~\cite{Bubanja:2023nrd}. 
The uncertainties labeled with \textit{data} at 68\% confidence level are obtained from the $q_s$ gap defined by $\chi^2=\chi^2_{min}+1$, and should be compared with the uncertainties labeled with \textit{data} in
Ref.~\cite{Bubanja:2023nrd}. Another source of uncertainties comes from the choice of the $p_T$ range. In Ref.~\cite{Bubanja:2023nrd} this is estimated by varying the numbers of bins in the fit. In this study we obtain it in the same way. For the last two $m_{DY}$ bins, it is not estimated due to the lack of statistics. 
We observe consistency with the result of 
 Ref.~\cite{Bubanja:2023nrd} in every $m_{DY}$ region.

In conclusion, we note  that the sensitivity of the $p_T(ll)$ shape to the intrinsic $k_T$ distribution lies in the shift of the spectrum from low $p_T$ to high $p_T$. This observation leads to the proposal of the $p_T(ll)$-ratio. We numerically test the sensitivity of the $p_T$-ratio to $q_s$ by comparing its statistical uncertainty with the same procedure on fine-binning $p_T$ structure. The result shows that the $p_T$-ratio has comparable sensitivity to the fine-binned $p_T$ shape.  

The methodology presented here can be used in
forthcoming experiments for TMD performance studies, as 
it circumvents the need to control large systematics in the low $p_T(ll)$ region. Unlike the extraction from a $p_T$ distribution that has been already binned, the $p_T$-ratio requires a dedicated study of the separation momentum $p_s$, affecting the sensitivity to TMD parameters, for example along the lines of the pseudo-data test carried out in this work.

\section*{Acknowledgments}

We thank the Rencontres de Moriond organizers for the pleasant atmosphere at the meeting and the 
invitation to present this work.  

\section*{References}
%\bibliography{moriond}

%%% manually generated bibliography
%\begin{thebibliography}{99}
%\bibitem{ja}C Jarlskog in {\em CP Violation}, ed. C Jarlskog
%(World Scientific, Singapore, 1988).
%\bibitem{ma}L. Maiani, \Journal{\PLB}{62}{183}{1976}.
%\bibitem{bu}J.D. Bjorken and I. Dunietz, \Journal{\PRD}{36}{2109}{1987}.
%\bibitem{bd}C.D. Buchanan {\it et al}, \Journal{\PRD}{45}{4088}{1992}.
%\end{thebibliography}

\end{document}